\documentclass{article}
\usepackage{spconf,amsmath,graphicx,hyperref}

\usepackage{booktabs}
\newcommand{\best}[1]{\textbf{#1}}

\usepackage{color}
\usepackage{threeparttable}
\usepackage[table]{xcolor}

\title{When Layer Selection Misleads Speech Depression Detection}

\name{Paula A. Perez-Toro$^{1,2\dagger}$ \qquad David Gimeno-Gómez$^{3\dagger}$ 
\qquad Daniel Rückert$^{1}$ 
\qquad Andreas Maier$^{2}$ 
\thanks{Corresponding author: paula.andrea.perez@fau.de.}}
\address{$^{1}$Chair for AI in Healthcare and Medicine, Technical University of Munich (TUM), Munich, Germany\\
$^{2}$Pattern Recognition Lab, Friedrich-Alexander-Universität Erlangen-Nürnberg, Erlangen, Germany\\
$^{3}$PRHLT research center, Universitat Politècnica de València (UPV), Valencia, Spain 
}

\begin{document}
\ninept
\maketitle
\begin{abstract}
Pretrained speech representations are increasingly used for depression detection, but selecting the best encoder layer on the same data used for evaluation biases reported performance. On DAIC-WOZ, with repeated nested cross-validation across five deep encoder families, naive best-of-25-layer probing inflates AUC by up to $0.09$. On shuffled labels over the \emph{real} latent representations, naive best-of-25 still reaches AUC $0.59$ versus $0.50$ under nested selection, and this bias grows with the number of probed layers and with smaller samples, isolating selection, not signal, as the cause. The effect replicates on a second clinical corpus and language (Androids, Italian). Under leakage-free selection, a fair comparison across representation families identifies a compact, task-aligned affect--prosody model as competitive with substantially more complex SSL, deep-learning, audio-LLM, and codec-based approaches, while using $\sim\!10^{3}\times$ fewer trainable downstream parameters.
Extending the analysis to individual PHQ-8 symptoms shows that the same selection bias persists at this finer-grained level. We argue nested model-selection protocols should be standard when probing encoder layer representations on small clinical speech datasets.
\end{abstract}
\section{Introduction}

Pretrained speech encoders are increasingly used as representations for clinical prediction. However, their high-dimensional and multi-layer representations introduce a methodological and rarely-audited model-selection issue: when the encoder layer is selected based on the same cross-validation folds used to report performance, the layer choice becomes part of the evaluation procedure. This matters in clinical speech tasks such as depression~\cite{Cummins2015}, where acoustic and prosodic characteristics have been shown to capture behavioural and symptomatic changes~\cite{gimeno2024reading,dumpala2024symptoms,rodriguez2024predicting,pereztoro2024longitudinal}. It is in this context of small clinical datasets, where many correlated layers can be explored relative to the number of participants, that such selection can produce optimistic estimates even when no predictive signal is present.

\begin{figure}[t]\centering
\includegraphics[width=1.0\columnwidth]{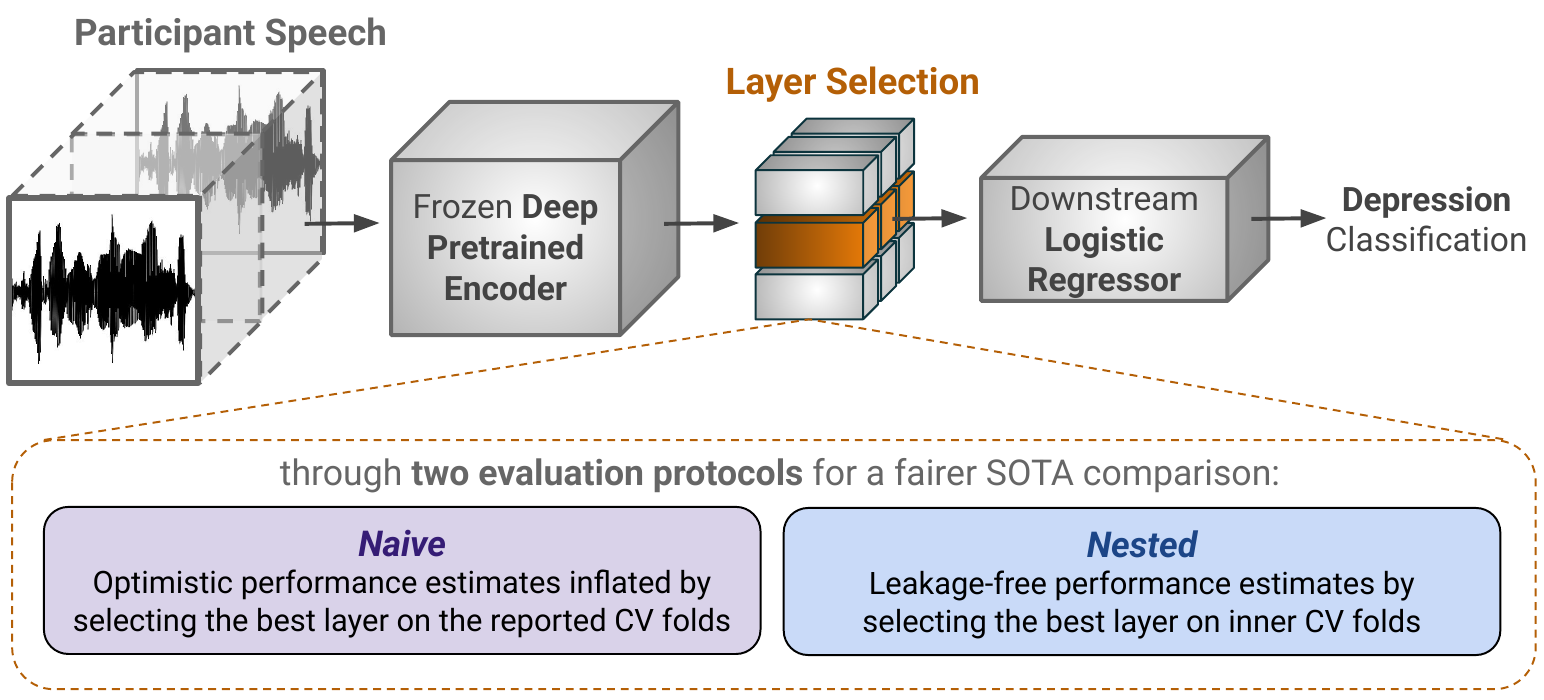}
\caption{Overview of the proposed framework for leakage-free comparison of pretrained speech representations. CV$=$Cross-Validation.}
\vspace{-0.45cm}
\label{fig:pipeline}
\end{figure}

The present study investigates this effect directly on the established DAIC-WOZ depression-detection benchmark~\cite{Gratch2014}, a dataset where, despite recent advances employing increasingly sophisticated representations, official test F1 performance remains at only $55$--$70\%$, while development performance routinely reaches $85$--$95\%$ with these approaches. Although this constitutes substantial progress over early systems based on handcrafted features with SVMs (F1 $0.35$--$0.50$ on AVEC-era data~\cite{Valstar2013}), as well as subsequent CNN- and attention-based models~\cite{ioannides2024daam,chen2022speechformer} and self-supervised speech representations~\cite{Baevski2020,Hsu2021}, the markedly gap in generalization remains poorly understood~\cite{Ravi2022step,wang2023nonuniform}. Our study therefore examines a complementary source of optimism that may contribute to this persistent gap.

Our contributions are threefold. (1) A \textbf{nested layer-selection audit} across five pretrained speech encoders, with pure-noise and label-permutation controls to quantify selection optimism and its dependence on candidate count, sample size, and corpus.\footnote{Source code and materials will be released upon acceptance.}
(2) A \textbf{fair comparison of representation families} under leakage-free evaluation, spanning interpretable, deep-learning, audio-LLM, and codec-based approaches, which identifies a compact \textbf{affect--prosody bottleneck} as competitive while using $\sim\!10^{3}\times$ fewer trainable downstream parameters.
(3) A \textbf{symptom-level analysis} extending the layer-selection audit and showing how naive selection consistently inflates AUC and can make pretrained representations appear more competitive with compact interpretable approaches.

\section{Method}
\label{sec:method}

\subsection{Data materials}
\label{sec:datasets}

\noindent\textbf{DAIC-WOZ}~\cite{Gratch2014} serves as our primary evaluation corpus, given its widespread adoption as an established benchmark for computational depression analysis. The corpus comprises semi-structured clinical interviews conducted by a Wizard-of-Oz animated interviewer in English, with depression assessed using the self-reported PHQ-8 instrument. We follow the official AVEC-2017 split~\cite{ringeval2017avec}, with a held-out test set of 47 participants. For cross-validation, we use the training ($n=107$) and development ($n=35$) sets, from which two subjects were excluded due to unavailable usable audio and/or PHQ-8 item labels, resulting in a total of 140 participants. The binary target label is defined by a PHQ-8 score $\geq 10$ indicating depression.

\vspace{0.15cm}
\noindent\textbf{Androids}~\cite{tao2023androids} represents our complementary evaluation setting in Italian, allowing us to assess whether our findings extend beyond the English language. The dataset includes multiple tasks and comprises 116 participants, of whom 62 are diagnosed with depression (directly provided as a binary target rather than using PHQ-8). For consistency with DAIC-WOZ, this study utilizes the spontaneous-speech interview task. Androids does not provide a separate held-out test set; therefore, we evaluate the full corpus using cross-validation.

\subsection{Speech representations}
\label{sec:layers}

To examine whether the layer-selection bias investigated in this study generalizes across different encoder families, we consider layer-wise representations from multiple \textit{frozen}, pretrained speech encoders of comparable scale ($\approx$0.3B parameters; exact checkpoints in released code), including the Large variants of \texttt{wav2vec2}~\cite{Baevski2020}, \texttt{data2vec}~\cite{baevski2022data2vec}, \texttt{WavLM}~\cite{chen2022wavlm}, \texttt{Whisper}~\cite{radford2023robust}, and \texttt{HuBERT}~\cite{Hsu2021}. For each encoder, we extract the input embedding and the outputs of all 24 transformber blocks, yielding 25 layers of 1024-dimensional representations. Each layer is then mean-pooled to obtain a single vector for each participant and layer.

\subsection{Evaluation and layer-selection protocols}
\label{sec:protocols}

As reflected in Fig.~\ref{fig:pipeline}, our proposed framework compares two evaluation protocol that differ specifically in how the best-performing encoder layer is selected: (1) \textbf{Naive.} The best-performing encoder layer is selected using the same cross-validation folds used to evaluate performance, potentially leading to optimistic performance estimates. (2) \textbf{Nested.} The encoder layer is selected using inner cross-validation within the training portion of each outer fold, separating layer selection from performance estimation.

Ultimately, this formulation can be viewed as a model-selection problem, suggesting that similar use of evaluation data in such optimization processes may introduce optimistic bias more broadly.

\subsection{Experimental setup}
\label{sec:setup}

\noindent\textbf{Cross-validation.} We perform $10$ repetitions of stratified, participant-disjoint $5$-fold cross-validation for both the \textit{naive} and \textit{nested} protocols. The key difference is that the \textit{nested} protocol further divides each outer training partition into $4$ inner folds for leakage-free encoder-layer selection. In all cases, the official test set of our primary DAIC-WOZ benchmark is kept separate from cross-validation and used only for final evaluation.

\vspace{0.15cm}
\noindent\textbf{Preprocessing.} For both datasets, we extract participant-only speech segments from the interview recordings, excluding interviewer interventions. We then apply variance thresholding and $z$-score standardization to the resulting representations, with all preprocessing parameters fitted using training data only.

\vspace{0.15cm}
\noindent\textbf{Downstream prediction.} In both protocols, the downstream predictor is the same fixed $L_2$-regularized logistic regression ($C=0.5$, class-balanced weights), with no hyperparameter search. Scaling, decision threshold, and Platt calibration are fitted using training data only, whereas each model is evaluated once on the held-out fold.

\vspace{0.15cm}
\noindent\textbf{Reference systems.} As a complementary reference, we additionally consider interpretable acoustic features that target characteristics associated with depressed speech and do not require model (layer) selection. These include (i) \textit{affect} (6-dim), consisting of the mean and standard deviation of arousal, dominance, and valence from the audEERING wav2vec-MSP emotion model~\cite{Wagner2023dawn}; (ii) \textit{timing} (25-dim) comprising pause, speech-rate, and response-latency statistics; and (iii) \textit{voice-quality} eGeMAPSv02 descriptors (88-dim)~\cite{Eyben2016}. Other reference systems comprise (iv) strong audio-language LLMs, represented by Qwen2-Audio~\cite{chu2024qwen2} through zero-shot probing, as well as (v) neural audio codecs, through Kyutai Mimi~\cite{kyutai2024moshi}, whose continuous representations consist of SEANet latents summarized by their mean and standard deviation and the discrete ones are summarized using RVQ-token statistics. For comparison, the same downstream predictor described above is used for all these reference systems.

\vspace{0.15cm}
\noindent\textbf{Statistical assessment.} To assess the reliability of the reported performance estimates and their deviation from chance, we use a 2000-sample participant-level bootstrap to quantify performance uncertainty and a shuffled-label null to provide a negative control.

\section{Results \& Discussions}

\subsection{Naive layer selection is optimistically biased}
\label{sec:audit}

\begin{figure*}[t]\centering
\includegraphics[width=0.88\linewidth]{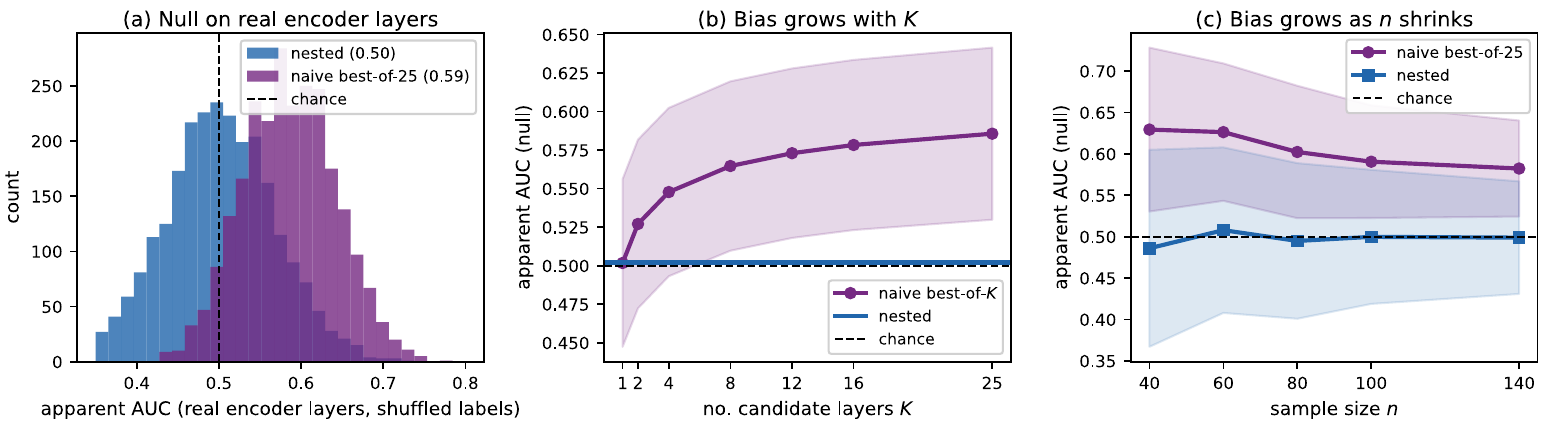}
\caption{Selection bias on real pretrained encoder layers with shuffled labels ($500$ permutations, $5$ encoders). (a)~Naive best-of-$25$ versus nested selection; (b)~null AUC as a function of the number of candidate layers $K$; (c)~null AUC as a function of sample size $n$.}
\vspace{-0.35cm}
\label{fig:null}
\end{figure*}

To directly investigate the effect of encoder layer selection on our primary benchmark, the well-established DAIC-WOZ corpus, Tab.~\ref{tab:audit} compares \textit{naive} and \textit{nested} selection across the considered encoder families. As can be observed, \textit{naively} selecting the best of 25 encoder layers on the same cross-validation folds used for evaluation consistently inflates AUC by $0.05$--$0.09$ points (mean $0.07$). The largest difference occurs for \texttt{wav2vec}, where AUC drops from $0.65$ with \textit{naive} selection to $0.56$ under \textit{nested} selection. Although the absolute difference may appear moderate, it substantially change the interpretation: the \textit{naive} estimate of $0.65$ suggests meaningful predictive performance, whereas the leakage-free estimate of $0.56$ is only slightly above chance, which could lead to misleading conclusions about system effectiveness in real-world deployment.
\begin{table}[!htpb]\centering\small
\caption{Layer-selection audit on the DAIC-WOZ cross-validation set using binary AUC (mean$\pm$std over $10$ CV repetitions). $\Delta$ denotes the optimistic mean difference between \emph{naive} and \emph{nested} performance. Higher-performing protocol highlighted in \textbf{bold}.}
\label{tab:audit}
\setlength{\tabcolsep}{21pt}
\resizebox{\linewidth}{!}{
\begin{tabular}{l@{\hspace{1cm}}cc@{\hspace{1cm}}c}
\toprule
\textbf{Encoder} & \textbf{\textit{naive}} & \textbf{\textit{nested}} & \textbf{$\Delta$} \\
\midrule
\texttt{wav2vec2} & \textbf{0.65{\scriptsize$\pm$0.02}} & 0.56{\scriptsize$\pm$0.03} & 0.09 \\
\texttt{data2vec} & \textbf{0.73{\scriptsize$\pm$0.03}} & 0.66{\scriptsize$\pm$0.05} & 0.07 \\
\texttt{WavLM}    & \textbf{0.63{\scriptsize$\pm$0.02}} & 0.56{\scriptsize$\pm$0.03} & 0.07 \\
\texttt{Whisper}  & \textbf{0.60{\scriptsize$\pm$0.03}} & 0.54{\scriptsize$\pm$0.03} & 0.06 \\
\texttt{HuBERT}   & \textbf{0.69{\scriptsize$\pm$0.02}} & 0.64{\scriptsize$\pm$0.03} & 0.05 \\
\bottomrule
\end{tabular}}
\end{table}

\vspace{0.15cm}
\noindent\textbf{Selection alone induces optimism.} We examine whether the observed optimism can arise purely from selection using $300$ signal-free simulations with $25$ Gaussian feature sets unrelated to the label. \textit{Naive} best-of-$25$ selection systematically yields above-chance apparent AUC ($0.64${\scriptsize$\pm0.03$}), exceeding $0.55$ in every trial, whereas nested selection remains at chance ($0.50${\scriptsize$\pm0.07$}). This demonstrates that selecting candidate layer representations on the full dataset can overstate the performance of pretrained speech encoders. The selected layer may also transfer poorly to held-out data, as for \texttt{data2vec}, whose naive AUC drops from $0.73$ in cross-validation to $0.62$ on the official test set.

To assess whether this selection bias persists in the actual encoder representations, we repeat the analysis with shuffled labels across the five speech encoders, as reflected in Fig.~\ref{fig:null}. Across $500$ permutations, preserving the inter-layer correlation, dimensionality, and participant structure, \textit{naive}
% best-of-$25$
selection still produces above-chance AUC (mean $0.59$), whereas \textit{nested} selection does not surpass chance (mean $0.50$, paired $\Delta$AUC
$+0.08$, $95\%$ CI $[0.08,0.09]$; Fig.~\ref{fig:null}a). The apparent AUC further increases monotonically with the number of candidate layers $K$ ($0.50\!\to\!0.59$ for
$K{=}1\!\to\!25$; Fig.~\ref{fig:null}b), consistent with the classical $\sigma\sqrt{2\log K}$ growth of the expected maximum for independent Gaussian estimates\footnote{This provides an upper bound here due to inter-layer correlation.}. Importantly, the optimism also increases as the sample size $n$ decreases ($\Delta$ from $0.08$ at $n{=}140$ to $0.14$ at $n{=}40$; Fig.~\ref{fig:null}c), while \textit{nested} remains close to chance. Together, the observed patterns highlight the impact of selection bias when exploring large numbers of candidate configurations in low-resource settings, as commonly encountered in clinical applications.

\vspace{0.15cm}
\noindent\textbf{The bias replicates across corpus and language.} Following the same shuffle-label setup, the Italian Androids Corpus (Tab.~\ref{tab:cross}) still reaches above-chance AUC ($0.58$) under \textit{naive} selection, while its \textit{nested} counterpart remains at chance ($0.50$). Notably, this occurs despite the encoders genuinely separating depression on the original no-shuffled labels (pooled AUC $0.96$ vs.\ $0.93$ for naive and nested selection, respectively), reinforcing that selection optimism would not specific to DAIC-WOZ, English, or low-signal regimes. It should be noted that we restrict Androids to this control because, in addition to its non-English speech making the emotion-encoder affect comparison an unfair cross-language test, it lacks the PHQ-8 item labels necessary for subsequent symptom-level analysis.

\begin{table}[!htpb]
\centering\small
\caption{Cross-validation evaluation on DAIC-WOZ and Androids under shuffled labels, pooled across the five pretrained encoders.}
\label{tab:cross}
\setlength{\tabcolsep}{6pt}
\resizebox{\linewidth}{!}{
\begin{tabular}{lccccccc}
\toprule
\textbf{Corpus} & \textbf{Lang} & \textbf{$n$} & & & \textbf{\textit{naive}} & \textbf{\textit{nested}} & \textbf{$\Delta$ [CI]} \\
\midrule
Androids & IT & 116 & & & 0.58 & 0.50 & $+$0.08 [0.08,0.08] \\
DAIC-WOZ & EN & 140 & & & 0.59 & 0.50 & $+$0.08 [0.08,0.09] \\
\bottomrule
\end{tabular}}
\end{table}

\vspace{-10pt}
\subsection{Fair comparison of speech representations}
\label{sec:main}

\begin{table*}[t]
\centering
\small
\caption{Comparison of reference systems on the DAIC-WOZ benchmark for the cross-validation and held-out test setups. CV ($n{=}140$) reports mean$\pm$std over $10$ repetitions; test ($n{=}47$) is a single, config-unstable held-out estimate ($\pm.13$ AUC). UAR and F1 use the OOF balanced-accuracy threshold; wF1 uses the OOF weighted-F1--optimal threshold. AUC is threshold-free. Pretrained speech encoders use nested layer selection; all other systems use fixed, pre-specified configurations. \best{Bold} = best and \underline{underline} = second-best per column.}
\label{tab:main}

\renewcommand{\arraystretch}{0.93}
\setlength{\tabcolsep}{16pt}

\begin{threeparttable}

\resizebox{0.95\linewidth}{!}{%
\begin{tabular}{@{}lcccc@{\hspace{20pt}}cccc@{}}
\toprule
& \multicolumn{4}{c}{\textbf{CV} ($n{=}140$)}
& \multicolumn{4}{c}{\textbf{Test} ($n{=}47$)} \\

\cmidrule(lr){2-5}
\cmidrule(l){6-9}

Model
& AUC & UAR & F1 & wF1
& AUC & UAR & F1 & wF1 \\

\midrule

\texttt{wav2vec2}
& 0.56{\scriptsize$\pm$0.03}
& 0.60{\scriptsize$\pm$0.02}
& 0.42{\scriptsize$\pm$0.09}
& 0.69{\scriptsize$\pm$0.02}
& 0.70
& 0.65
& 0.53
& 0.67 \\

\texttt{data2vec}
& 0.66{\scriptsize$\pm$0.05}
& \underline{0.65{\scriptsize$\pm$0.04}}
& \underline{0.51{\scriptsize$\pm$0.04}}
& \underline{0.70{\scriptsize$\pm$0.03}}
& 0.54
& 0.52
& 0.36
& 0.60 \\

\texttt{WavLM}
& 0.56{\scriptsize$\pm$0.03}
& 0.58{\scriptsize$\pm$0.02}
& 0.44{\scriptsize$\pm$0.04}
& 0.64{\scriptsize$\pm$0.03}
& 0.73
& 0.63
& 0.51
& 0.74 \\

\texttt{Whisper}
& 0.54{\scriptsize$\pm$0.03}
& 0.56{\scriptsize$\pm$0.03}
& 0.41{\scriptsize$\pm$0.05}
& 0.65{\scriptsize$\pm$0.03}
& 0.77
& \textbf{0.77}
& \textbf{0.67}
& 0.67 \\

\texttt{HuBERT}
& 0.64{\scriptsize$\pm$0.03}
& 0.64{\scriptsize$\pm$0.03}
& 0.49{\scriptsize$\pm$0.05}
& \underline{0.70{\scriptsize$\pm$0.03}}
& 0.65
& 0.67
& \underline{0.55}
& 0.67 \\

\addlinespace[3pt]

\rowcolor{gray!15}
\multicolumn{9}{l}{\textit{Interpretable acoustic features}} \\

Affect (6-d)
& 0.65{\scriptsize$\pm$0.02}
& 0.64{\scriptsize$\pm$0.02}
& \underline{0.51{\scriptsize$\pm$0.03}}
& 0.69{\scriptsize$\pm$0.01}
& \best{0.89}
& 0.65
& \underline{0.55}
& \best{0.87} \\

Affect+Timing (31)
& \underline{0.67{\scriptsize$\pm$0.02}}
& \underline{0.65{\scriptsize$\pm$0.02}}
& \underline{0.51{\scriptsize$\pm$0.04}}
& \underline{0.70{\scriptsize$\pm$0.01}}
& \underline{0.85}
& 0.66
& \underline{0.55}
& \underline{0.77} \\

\textbf{Affect+Timing+eGeMAPS (119)}
& \best{0.75{\scriptsize$\pm$0.03}}
& \best{0.72{\scriptsize$\pm$0.03}}
& \best{0.59{\scriptsize$\pm$0.03}}
& \best{0.74{\scriptsize$\pm$0.01}}
& 0.72
& 0.65
& 0.54
& 0.67 \\

\addlinespace[3pt]

\rowcolor{gray!15}
\multicolumn{9}{l}{\textit{Zero-shot audio-language LLM}} \\

Qwen2-Audio\textsuperscript{$\dagger$}
& 0.60
& 0.59
& 0.42
& 0.68
& 0.84
& \underline{0.76}
& \textbf{0.67}
& 0.76 \\

\addlinespace[3pt]

\rowcolor{gray!15}
\multicolumn{9}{l}{\textit{Neural audio codec representations}} \\

Mimi codec (continuous, 1024)
& 0.47{\scriptsize$\pm$0.03}
& 0.55{\scriptsize$\pm$0.02}
& 0.43{\scriptsize$\pm$0.08}
& 0.61{\scriptsize$\pm$0.02}
& 0.55
& 0.52
& 0.26
& 0.61 \\

Mimi codec (discrete, 128)
& 0.58{\scriptsize$\pm$0.02}
& 0.61{\scriptsize$\pm$0.02}
& 0.46{\scriptsize$\pm$0.03}
& 0.67{\scriptsize$\pm$0.01}
& 0.49
& 0.51
& 0.38
& 0.66 \\

\bottomrule
\end{tabular}%
}

\begin{tablenotes}
\footnotesize
\item[\textsuperscript{$\dagger$}]
Zero-shot evaluation uses yes/no output logits and is deterministic; therefore, no standard deviation is reported.
\end{tablenotes}

\end{threeparttable}
\end{table*}

Having established the optimism introduced by naive layer selection, we next compare the leakage-free nested evaluation protocol with reference systems spanning a broad range of speech representations. Tab.~\ref{tab:main} thus aims to provide a fairer assessment of the relative performance achieved by these approaches in the field.

According to the results, \textit{affect} alone attains a nested-CV AUC $0.65${\scriptsize$\pm0.02$}, whereas its combination with \textit{timing} gives only a marginal increase. It is the addition of the eGeMAPSv02 features
% to this combination
that markedly raises performance to $\mathbf{0.75}${\scriptsize$\mathbf{\pm0.03}$} ($95\%$ CI $[0.68,0.84]$; shuffled-label null $0.52${\scriptsize$\pm0.07$}). This $0.65\!\to\!0.75$ gain suggests that \textit{voice-quality} features, including H1-H2, provide complementary information beyond \textit{affect}, while \textit{timing}, which is near-null on binary AUC, carries symptom-specific signal (Sec.~\ref{sec:symptom}).

Turning to the deep speech representations under \emph{nested} layer selection, the best pretrained encoder, \texttt{data2vec}, reaches an AUC of $0.66${\scriptsize$\pm0.05$}, comparable to \textit{affect} alone but below the best-performing interpretable setting. A general audio-language model, \texttt{Qwen2-Audio-7B}, likewise does not exceed the interpretable representations: its training-free zero-shot evaluation reaches an AUC of $0.60$, while an additional frozen-encoder probe using its latent embeddings even performed worse ($0.56${\scriptsize$\pm0.03$}). Finally, Kyutai Mimi performs weakest among the considered representation families, with AUCs of $0.47${\scriptsize$\pm0.03$} for continuous SEANet latents and $0.58${\scriptsize$\pm0.02$} for discrete RVQ-token statistics. The proposed $119$-dim acoustic representation \texttt{Affect+Timing+eGeMAPS} therefore achieves the highest cross-validation AUC among the considered systems while requiring only $\approx\!\!120$ trainable parameters.

\vspace{0.15cm}
\noindent\textbf{The held-out test scores are high but ranking-unstable.} On the $47$-speaker test set, the fixed \textit{affect}+logistic model reaches an AUC of $\mathbf{0.89}$ $[0.76,0.98]$, illustrating that a high individual test estimate can coexist with substantial ranking instability. The CV ranking does not carry over to the test set: the CV-best \texttt{Affect+Timing+eGeMAPS} drops from $0.75$ to $0.72$, while \texttt{Whisper}, which ranks last among the pretrained deep encoders in CV ($0.54$), reaches $0.77$ on test. Similarly, \texttt{Qwen2-Audio} (zero-shot) increases from $0.60$ in CV to $0.84$ on test, although the same output logits varied substantially ($0.89$/$0.76$) by classifier configuration alone. Overall, CV and test rankings show only weak agreement across all systems (Spearman $\rho{=}0.25$) and are strongly anti-correlated among the five pretrained encoders ($\rho{=}-0.97$). These rank inversions highlight the instability of model comparisons on a small held-out test set and motivate using repeated nested CV as the primary basis for ranking systems.

\vspace{0.15cm}
\noindent\textbf{Complexity, robustness, and what does not help.} To study whether increased model complexity provides some benefit, we additionally evaluated four deep heads (transformer, cross-attention, multitask, and supervised-contrastive) over encoder-layer tokens. None improved upon the simpler representations: the best configuration, cross-attention with \texttt{data2vec}, reached only $0.60${\scriptsize$\pm0.06$} under $10$-repeat cross-validation. These deep-head experiments also exhibited high-variance training, limiting the reliability of single-run estimates (per-architecture results in released code).

Beyond the deep architectures above, additional feature ablations further show what does not improve performance. \textit{Affect dynamics}, represented by $38$-dimensional trajectory descriptors capturing blunting, instability, inertia, and coupling, consistently reduce performance, with a decrease from $0.75$ to $0.70$ when appended to the existing feature sets. Other extensions, including multitask symptom learning and SMOTE~\cite{chawla2002smote}, likewise do not improve performance. An exhaustive sweep over all $15$ feature subsets further confirmed that \texttt{Affect+Timing+eGeMAPS} is uniquely CV-optimal (AUC $0.75$; next-best \textit{Affect+eGeMAPS}, AUC $0.69$).

In terms of robustness, the result does not appear to be explained by sex: although the features predict sex with AUC $1.0$ and sex prevalence differs between depressed and non-depressed participants ($0.39$ vs.\ $0.23$), depression detection remains strong within both females and males (AUC $0.73$/$0.77$, respectively) and is unchanged after residualizing the features for sex (AUC $0.76\!\to\!0.75$).

\subsection{Symptom profiling: validated routing, coherent severity}
\label{sec:symptom}

\begin{figure}[!htpb]\centering
\includegraphics[width=0.85\columnwidth]{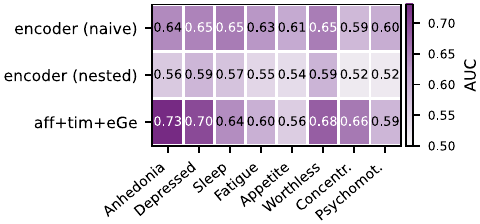}
\caption{Per-symptom AUC ($3$-rep CV, encoders pooled). \emph{naive} vs \emph{nested} layer
selection: the bias replicates on all $8$ PHQ-8 items; the selection-free \texttt{aff+tim+eGe}
exceeds \emph{nested} on every one.}
\vspace{-0.5cm}
\label{fig:sympt}
\end{figure}

As Fig.~\ref{fig:sympt} depicts, our downstream models decompose into a PHQ-8 symptom profile at no additional cost, allowing us to extend the layer-selection audit to a finer-grained level. Pooled across the five encoders, \textit{naive} selection inflates AUC for all eight symptoms (mean $0.63$ vs.\ $0.56$ for \textit{nested}). While the leakage-free \textit{nested} protocol performs worse than the \texttt{Affect+Timing+eGeMAPS} reference system on all eight symptoms (mean AUC gap $0.09$, range $[0.02,0.17]$), \textit{naive} selection exceeds the reference system (mean AUC gap $0.04$, range $[0.01,0.09]$) on $4/8$ symptoms (sleep, fatigue, appetite, and psychomotor). This contrast further supports our central finding that evaluation leakage can alter the apparent competitiveness of pretrained representations relative to more efficient, interpretable approaches, even across multiple levels of modeling.

Using weighted-F1 with score${>}0$ binarization, our symptom predictions also match or exceed the CNN/LSTM approach of~\cite{rodriguez2024predicting} ($\mathbf{0.65}$ vs.\ $0.59$; better on $6/8$ symptoms). Summing per-item cumulative-ordinal expectations ($\hat{y}_s=\sum_k P(y_s\!\ge\!k)$) yields severity with $4$-class $\kappa\,0.46${\scriptsize$\pm0.03$} (vs.\ $0.34$ for direct regression); probabilities
are over-confident (ECE $0.15${\scriptsize$\pm0.02$}), improved by Platt scaling to
$0.09${\scriptsize$\pm0.02$}. This per-symptom profile, coherent severity, and calibrated
probabilities are what a screening tool needs but a black-box latent embedding score lacks.
\vspace{-0.3cm}

\section{Conclusions \& Limitations}
\label{sec:limit_conclusions}

Across two clinical speech corpora and languages, selecting pretrained encoder layers on the same folds used for reporting produces systematic optimism that grows with the number of candidates and decreases with sample size. Nested selection removes this bias, can change the ranking of representations, and should therefore be standard practice when probing encoder layers on small clinical speech datasets. Under this leakage-free evaluation, a compact task-aligned affect--prosody model emerges as a competitive alternative to substantially more complex deep-learning, audio-LLM, and codec models, while additionally providing interpretable and calibrated symptom-level outputs.

Several limitations bound these conclusions. While the selection-bias finding replicates on Androids ($n{=}116/62$), the broader representation comparison and symptom analyses remain single-corpus results (DAIC-WOZ: $n{=}140/47$). Moreover, the wav2vec-MSP affect encoder is English-trained. Although it transfers well to Italian (AUC $0.87$), it is outperformed there by multilingual representations; thus, conclusions from the reference-system comparison are specific to DAIC-WOZ and to \emph{task-aligned compression}, rather than classical signal processing. Finally, our proof-of-concept focuses on encoder-layer selection. Although the differences across encoder families (Tab.~\ref{tab:audit}) suggest broader implications for architecture exploration, whether reuse of evaluation data in related optimization procedures produces similar optimism remains an open question.

\section{Acknowledgments}

The work of Gimeno-Gómez was partially supported by the PROMETEO 2024 program (project LightVED, CIPROM/2023/17), funded by Generalitat Valenciana (GVA).

\section{Compliance with Ethical Standards}
This study is a secondary analysis of two pre-existing, de-identified public speech corpora,
DAIC-WOZ~\cite{Gratch2014} and Androids~\cite{tao2023androids}; no new participants were recruited
or data collected. Both were used under their data-use terms, and their original collections were
approved by the respective institutional review boards/ethics committees.

\bibliographystyle{IEEEbib}
\bibliography{main}

@inproceedings{chen2022speechformer,
  author = {Chen, W. and Xing, X. and Xu, X. and Pang, J. and Du, L.},
  title = {{SpeechFormer: A Hierarchical Efficient Framework Incorporating the Characteristics of Speech}},
  booktitle = {Proc. of Interspeech},
  pages = {346--350},
  year = {2022},
  doi = {10.21437/Interspeech.2022-74},
  url = {https://arxiv.org/abs/2203.03812}
}

@inproceedings{wang2023nonuniform,
  title={{Non-uniform Speaker Disentanglement For Depression Detection From Raw Speech Signals}},
  author={Wang, J. and Ravi, V. and Alwan, A.},
  booktitle={Proc. of Interspeech},
  pages={2343--2347},
  year={2023},
  doi={https://arxiv.org/abs/2306.01861}
}

@inproceedings{gimeno2024reading,
  title={{Reading Between the Frames: Multi-Modal Depression Detection in Videos from Non-Verbal Cues}},
  author={Gimeno-G{\'o}mez, D. and Bucur, A.-M. and Cosma, A. and Mart{\'\i}nez-Hinarejos, C.-D. and Rosso, P.},
  booktitle={Proc. of ECIR},
  pages={191--209},
  year={2024},
  organization={Springer}
}

@article{ioannides2024daam,
  title={{Density Adaptive Attention-based Speech Network: Enhancing Feature Understanding for Mental Health Disorders}},
  author={Ioannides, G. and Kieback, A. and Chadha, A. and Elkins, A.},
  journal={arXiv preprint arXiv:2409.00391},
  year={2024}
}

@inproceedings{Ravi2022step,
  title={{A Step Towards Preserving Speakers' Identity While Detecting Depression Via Speaker Disentanglement}},
  author={Ravi, V. and Wang, J. and Flint, J. and Alwan, A.},
  booktitle={Proc. of Interspeech},
  pages={3338--3342},
  year={2022},
  url={https://pmc.ncbi.nlm.nih.gov/articles/PMC9635494/}
}

@article{Cummins2015,
  author = {Cummins, N. and Scherer, S. and Krajewski, J. and Schnieder, S. and Epps, J. and Quatieri, T. F.},
  title = {A review of depression and suicide risk assessment using speech analysis},
  journal = {Speech Communication},
  volume = {71},
  pages = {10--49},
  year = {2015},
  doi = {10.1016/j.specom.2015.03.004}
}

@inproceedings{Gratch2014,
  author = {Gratch, J. and Artstein, R. and Lucas, G. M. and Stratou, G. and Scherer, S. and Nazarian, A. and Wood, R. and Boberg, J. and DeVault, D. and Marsella, S. and Traum, D. and Rizzo, S. and Morency, L.-P},
  title = {{The Distress Analysis Interview Corpus of human and computer interviews}},
  booktitle = {Proc. of LREC},
  pages = {3123--3128},
  year = {2014}
}

@inproceedings{Baevski2020,
  author = {Baevski, A. and Zhou, Y. and Mohamed, A. and Auli, M.},
  title = {{wav2vec 2.0: A Framework for Self-Supervised Learning of Speech Representations}},
  booktitle = {Proc. of NeurIPS},
  volume = {33},
  pages = {12449--12460},
  year = {2020}
}

@inproceedings{baevski2022data2vec,
  title={{data2vec: A general framework for self-supervised learning in speech, vision and language}},
  author={Baevski, A. and Hsu, W.-N.and Xu, Q. and Babu, A. and Gu, J. and Auli, M.},
  booktitle={Proc. of ICML},
  pages={1298--1312},
  year={2022},
  organization={PMLR}
}

@article{Hsu2021,
  author = {Hsu, W.-N. and Bolte, B. and Tsai, Y-H. and Lakhotia, K. and Salakhutdinov, R. and Mohamed, A.},
  title = {{HuBERT}: Self-supervised speech representation learning by masked prediction of hidden units},
  journal = {IEEE/ACM TASLP},
  volume = {29},
  pages = {3451--3460},
  year = {2021},
  doi = {10.1109/TASLP.2021.3122291}
}

@article{chen2022wavlm,
  title={{WavLM: Large-Scale Self-Supervised Pre-Training for Full Stack Speech Processing}},
  author={Chen, S. and Wang, C. and Chen, Z. and Wu, Y. and Liu, S. and Chen, Z. and Li, J. and Kanda, N. and Yoshioka, Tm and Xiao, X. and others},
  journal={IEEE JSTSP},
  volume={16},
  number={6},
  pages={1505--1518},
  year={2022},
  publisher={IEEE}
}

@inproceedings{radford2023robust,
  title={{Robust Speech Recognition via Large-Scale Weak Supervision}},
  author={Radford, A. and Kim, J. and Xu, T. and Brockman, G. and McLeavey, C. and Sutskever, I.},
  booktitle={Proc. of ICML},
  pages={28492--28518},
  year={2023},
  organization={PMLR}
}

@article{chawla2002smote,
  title={{SMOTE: Synthetic Minority Over-Sampling Technique}},
  author={Chawla, N. V. and Bowyer, K. W. and Hall, L. O. and Kegelmeyer, W. P.},
  journal={Journal of Artificial Intelligence Research},
  volume={16},
  pages={321--357},
  year={2002}
}

@inproceedings{Valstar2013,
  author = {Valstar, Miñ. and Schuller, B. and Smith, K. and Eyben, F. and Jiang, B. and Bilakhia, S. and Schnieder, S. and Cowie, R. and Pantic, M.},
  title = {{AVEC} 2013: The continuous audio/visual emotion and depression recognition challenge},
  booktitle = {Proc. of ACM International AVEC Workshop},
  pages = {3--10},
  year = {2013},
  doi = {10.1145/2512530.2512533}
}

@inproceedings{ringeval2017avec,
  title = {{AVEC 2017: Real-life Depression, and Affect Recognition Workshop and Challenge}},
  author = {Ringeval, F. and Schuller, B. and Valstar, M. and Gratch, J. and Cowie, R. and Scherer, S. and Mozgai, S. and Cummins, N. and Schmitt, M. and Pantic, M.},
  year = {2017},
  booktitle = {Proc. of ACM International AVEC Workshop},
  pages = {3--9},
}

@article{Eyben2016,
  author = {Eyben, F. and Scherer, K. R. and Schuller, B. W. and Sundberg, J. and Andr{\'e}, E. and Busso, C. and Devillers, L. Y. and Epps, J. and Laukka, P. and Narayanan, S. S. and Truong, K. P.},
  title = {The {Geneva} Minimalistic Acoustic Parameter Set ({GeMAPS}) for voice research and affective computing},
  journal = {IEEE Transactions on Affective Computing},
  volume = {7},
  number = {2},
  pages = {190--202},
  year = {2016},
  doi = {10.1109/TAFFC.2015.2457417}
}

@inproceedings{dumpala2024symptoms,
  title     = {{Self-Supervised Embeddings for Detecting Individual Symptoms of Depression}},
  author    = {Dumpala, S. H. and Dikaios, K. and Nunes, A. and Rudzicz, F. and Uher, R. and Oore, S.},
  booktitle = {Proc. of Interspeech},
  pages     = {1450--1454},
  year      = {2024}
}

@article{Wagner2023dawn,
  title   = {{Dawn of the Transformer Era in Speech Emotion Recognition: Closing the Valence Gap}},
  author  = {Wagner, J. and Triantafyllopoulos, A. and Wierstorf, H. and Schmitt, M. and Burkhardt, F. and Eyben, F. and Schuller, B. W.},
  journal = {IEEE TPAMI},
  volume  = {45}, number = {9}, pages = {10745--10759}, year = {2023}
}

@article{chu2024qwen2,
  title={{Qwen2-Audio Technical Report}},
  author={Chu, Y. and Xu, J. and Yang, Q. and Wei, H. and Wei, X. and Guo, Z. and Leng, Y. and Lv, Y. and He, J. and Lin, J. and Zhou, C. and Zhou, J.},
  journal={arXiv preprint arXiv:2407.10759},
  year={2024}
}

@techreport{kyutai2024moshi,
    author = {A. D\'efossez and L. Mazar\'e and M. Orsini and A. Royer and P. P\'erez and H. J\'egou and E. Grave and N. Zeghidour},
    title = {{Moshi: A Speech-Text Foundation Model for Real-Time Dialogue}},
    institution = {Kyutai},
    year={2024},
    month={September},
    url={http://kyutai.org/Moshi.pdf},
}

@inproceedings{tao2023androids,
  title={{The Androids Corpus: A New Publicly Available Benchmark for Speech Based Depression Detection}},
  author={Tao, F. and Esposito, A. and Vinciarelli, A.},
  booktitle={Proc. of Interspeech},
  pages={4149--4153},
  year={2023}
}

@article{rodriguez2024predicting,
  title={{Predicting Individual Depression Symptoms from Acoustic Features During Speech}},
  author={Rodriguez, S. and Dumpala, S. H. and Dikaios, K. and Rempel, S. and Uher, R. and Oore, S.},
  journal={arXiv preprint arXiv:2406.16000},
  year={2024}
}

@inproceedings{pereztoro2024longitudinal,
  author={P{\'e}rez-Toro, P. A. and Dineley, J. and Kaczkowska, A. and Conde, P. and Zhang, Y. and Matcham, F. and Siddi, S. and Haro, J. M. and Bruce, S. and Wykes, T. and Bail{\'o}n, R. and Vairavan, S. and Dobson, R. J.B. and Maier, A. and N{\"o}th, E. and Orozco-Arroyave, J. R. and Narayan, V. A. and Cummins, N.},
  title={{Longitudinal Modeling of Depression Shifts Using Speech and Language}},
  booktitle={Proc. of ICASSP},
  pages={12021--12025},
  year={2024},
  doi={10.1109/ICASSP48485.2024.10447195}
}

\end{document}